
%
%
%
%
\magnification=1100 \tolerance=1000
\baselineskip=13pt plus 2pt minus 1pt \parskip=5pt plus 2pt minus 1pt
\font\flrm=cmr10 at 14.4truept \font\fvlrm=cmr10 at 20.74truept
\font\frm=cmr10 at 12truept \font\flbf=cmr10 at 14.4truept
\global\newcount\eqn  \global\eqn=0
\global\newcount\sec \global\sec=0
\global\newcount\ftno \global\ftno=0
\global\newcount\prg \global\prg=0
\def\neq{\global\advance\eqn by1 \eqno(\the\sec . \the\eqn)}
\def\section#1{\global\advance\sec by1 \bigskip \bigskip
\goodbreak
\centerline{{\flrm \the\sec.- #1}}\par\nobreak\nobreak \global\eqn=0
\global\prg=0}
\def\parag#1{\global\advance\prg by1 \bigskip \goodbreak
\centerline{\frm \the\sec.{\oldstyle \the\prg} - #1}\par\nobreak}
\def\label#1{\xdef#1{\hbox{(\the\sec.\the\eqn)}}}
\def\ftnote#1{\global\advance\ftno by1 \footnote{$^{\the\ftno}$}{#1}}
\def\frac#1#2{{#1 \over #2}}
\def\ph{\varphi}
\def\eps{\varepsilon}
\def\norm#1{\,\vert #1\vert\,}
\def\corr#1{\ \langle #1\rangle\ }
\def\sqr#1#2#3{{\vcenter{\hrule height.#3pt \hbox{\vrule width.#3pt
height#1pt \kern#1pt \vrule width.#2pt} \hrule height.#2pt}}}
\def\sq{{\mathchoice\sqr652\sqr552\sqr{2.1}31\sqr{1.5}31}\hskip 1.5pt}
%
%
\global\newcount\refno \global\refno=0
\newwrite\rfile
\def\ref#1#2{\global\advance\refno by1 \xdef#1{[\the\refno]}
\ifnum\refno=1\immediate\openout\rfile=refs.aux\fi
\immediate\write\rfile{\noexpand \item{[\the\refno]}{#2}}}


\ref\AZ{A.B. Zamolodchikov, JETP Lett. {\bf 43}, 730 (1986) ;
Sov. J. Nucl. Phys. {\bf 46}, 1090 (1987).}

\ref\C{J.L. Cardy, Phys. Rev. Lett. {\bf 60} (1988) 2709.}

\ref\F{D. Friedan, unpublished.}

\ref\CFL{A. Cappelli, D. Friedan, J.I. Latorre, Nucl. Phys.
{\bf B352}, 616 (1991).}

\ref\OJS{H. Osborn, Phys. Lett. {\bf B222}, 97 (1989) ;
I. Jack and H. Osborn, Nucl. Phys. {\bf B343}, 647 (1990) ;
G. Shore, Phys. Lett. {\bf B253}, 380 (1991) ; {\bf B256}, 407
(1991).}

\ref\CLV{A.Cappelli, J.I.Latorre and X.Vilas\'\i s-Cardona,
Nucl.Phys.{\bf B376}, 510 (1992).}

\ref\KZ{V.G.Knizhnik and A.B.Zamolodchikov, Nucl.Phys. {\bf
B247}, 83
(1984).}

\ref\IZ{C. Itzykson, J.B. Zuber, ``Quantum Field Theory",
McGraw-Hill, New York (1980).}

\ref\FLV{D.Z.Freedman, J.I.Latorre and X.Vilas\'\i s-Cardona,
Mod.Phys.Lett. {\bf A6}, 531 (1991).}

\ref\FS{Y.Frishman, A.Schwimmer, T.Banks, S.Yankielowicz, Nucl.Phys.
{\bf B177}, 157 (1981).}

\ref\OP{H. Osborn, A. Petkos, ``Implications of conformal
invariance in
field theories for general dimensions", DAMTP-preprint 93-31
(1993).}

\ref\W{E.Witten, Commun.Math.Phys. {\bf 92}, 455 (1984).}

\ref\G{P. Ginsparg in ``Fields, strings and critical Phenomena",
{\sl Proceedings of the Les Houches summer school 1988}, ed.
E.Br\'ezin,
J.Zinn-Justin, North Holland, Amsterdam (1990).}

\line{UB-ECM-PF 94/6\hfill April 1994}
\leftline{hep-th/9404150}
\vskip 1truecm
\centerline{\fvlrm Renormalisation Group Flows}
\bigskip
\centerline{\fvlrm and Conserved Vector Currents}
\vskip 1truecm
\centerline{\flbf Xavier Vilas\'\i s-Cardona}\bigskip
\centerline{D.E.C.M. and I.F.A.E.}
\centerline{Facultat de F\'\i sica, Universitat de Barcelona}
\centerline{Diagonal 647, E-08028 Barcelona, Catalonia, Spain}
\centerline{e-mail : {\tt druida@ecm.ub.es}}
\vskip 1truecm
\centerline{\flbf Abstract}
\bigskip
{\narrower\frm
Irreversibility of RG flows in two dimensions is shown using
conserved vector currents. Out of a conserved vector current, a
quantity
decreasing along the RG flow is built up such that it is
stationary at
fixed points where it coincides with the constant coefficient of
the two
current correlation function. For Wess-Zumino-Novikov-Witten
models this
constant coefficient is the level of their associated affine Lie
algebra. Extensions to higher dimensions using the
spectral
decomposition of the two current correlation function are studied.\par}
\vskip 1truecm

\section{Introduction}
Zamolodchikov has established
the irreversibility of Renormalisation Group flows in two
dimensions , through his celebrated $c$-theorem \AZ.
The theorem is proved by
constructing a quantity that is monotonically decreasing along
the RG
flow and is stationary at fixed points where it coincides with
the
central charge of the corresponding Conformal Field Theory. Three
are the basic ingredients to the demonstration : Lorentz invariance,
unitarity and existence and conservation all along the RG flow of the
stress tensor. The difference between the UV and IR central
charges is shown to be computable
from the correlation function of two traces of the stress tensor
using a very simple sum rule \C. An alternative proof of the
$c$-theorem which uses
the spectral decomposition of the correlation function
of precisely two stress
tensors has been given in references \F\ and \CFL. This is a very
convenient
picture to understand the $c$-theorem in terms of loss of degrees
of freedom. In effect, the central charge is a measure of the degrees of
freedom in the theory. When the UV CFT is relevantly perturbed, some
will become massive and decouple in the IR
limit. This feature is explicitely seen on the
spectral representation of the two stress tensor correlation function.
Recall that the stress tensor is the quantity to which all degrees of
freedom couple. In unitary theories, where all degrees of freedom
add positively to their counting, the central charge will
effectively decrease.
By integrating from the spectral density those degrees of freedom that
have become massive, one can
quantify how many are lost along the RG flow. In this
manner a number of sum rules can be build up, Cardy's \C\ among them.
There have been several attempts to enlarge the scope of the
$c$-theorem beyond two dimensions, though no general conclusive
result has been attained so far \OJS\ \CFL\ \CLV.

The purpose of this paper is to elaborate on a similar result for
two dimensional theories with conserved vector currents \F.
In effect, one can mimic both Zamolodchikov's and
spectral proof of the $c$-theorem substituting the stress tensor
by a conserved vector current. One constructs in this manner a quantity
monotonically decreasing along the RG flow and stationary at fixed
points where it coincides with the constant coefficient of the
two current correlation function. Whenever vector currents
satisfy a Kac-Moody algebra this constant coefficient
corresponds to the level of its central extension \KZ.
In this sense, we shall colloquially talk about the $k$-theorem.
Originally, Friedan proved the $k$-theorem using the spectral
representation technique \F.
The basic conditions required to formulate and prove it
are mostly the same one needs for the $c$-theorem, namely, the
existence of a conserved vector current along the RG flow, Lorentz
invariance and unitarity. Let us insist that the only requirement
made on the current is its conservation. One should not be mislead by
the particular case of
fermion theories where the two point function of the vector
current encodes information about the chiral anomaly.
The $k$-theorem is a general statement on conserved currents
and, as such, has no relation with the chiral anomaly.

Just like the $c$-theorem, the $k$-theorem portrays
the loss along the RG of the degrees of freedom coupling to the
vector current. These are accounted at CFT by the constant
coefficient of the two current correlation function.
In contradistinction to the $c$-theorem case,
it may happen that none of the degrees of freedom
coupling to the current becomes massive under a certain relevant
perturbation. Then, the constant coefficient remains the
same all along the flow. The quantification of the change in the
constant coefficient is achieved through a number of sum rules,
just like in the $c$-theorem case.

In order to enlarge the scope of the theorem to higher
dimensions, the
spectral proof seems to provide a natural framework. This is because the
formalism is set up with no reference to the dimension of
space time. Besides, the $k$-theorem case looks a
priori simpler than the $c$-theorem one. In effect, when looking
for an
an extension of the spectral proof for the $c$-theorem, one has
to consider two spin structures for the decomposition of the two
stress tensor correlation function while in two dimensions, only
one structure is present. Then, the information encoded in the
central charge is split into the coefficients of each structure and
gets so mistified. For conserved vector currents, only
one structure occurs in two as well as in more dimensions. This is
because the two possible possible structures one can build for the two
point function with
one momentum and the metric tensor are constrained to one by the current
Ward Identity.
Although this is a simplification with respect to the $c$-theorem
shall see that it does not keep up expectations.

The paper unfolds as follows.
In section 2, the irreversibility of RG flows for
two dimensional theories with conserved vector currents is proved
using
techniques similar to Zamolodchikov's in his original proof of
the $c$-theorem. From that, the equivalent of the ``Cardy
sum rule" is derived.
In section 3, the spectral decomposition of the two current
correlation function
is studied and the original proof by Friedan \F\ of the $k$-theorem
for two dimensional theories is reproduced.
The previous sum rule will then be recovered in
this framework, and new sum rules will be obtained. The features
described in sections 2 and 3 are then
illustrated in section 4 by a simple example, two dimensional
free
massive fermions. Finally, the naive extension of the $k$-theorem
using
the spectral representation is considered in section 5, computing
explicitely as a particular case
the spectral density of gauge currents for massive
bosons and fermions.

\section{RG flows in two dimensions and conserved vector
currents}
Let us start by giving a proof of the irreversibility of RG flows
for
theories with conserved vector currents in two dimensions which
mimics
Zamolodchikov's proof of the $c$-theorem \AZ, since this is the
historical
standard and the way of reasoning the reader might be
more familiar with.

Consider a two-dimensional Quantum Field Theory which is
Euclidean invariant and reflection positive (reflection
positivity amounts to the Euclidean equivalent of unitarity at the
level of Green functions). Let $J_\alpha (x)$ be a vector conserved
current in the theory, at classical and quantum level.
Let us introduce complex coordinates,
$ z=x_0+{\rm i}x_1,$ $\overline{z}=x_0-{\rm i}x_1$
and the following notation,
$$ \eqalign{J(z)&=J_z(z,\overline{z}),\cr
\overline{J}(z)&=J_{\overline{z}}(z,\overline{z}).\cr}$$
Consider the two current correlation function.
Euclidean invariance arguments and the absence of anomalous
dimensions for a conserved current allow us to write
$$ \eqalign{
\corr{J(z)J(w)}&={R(\tau)\over(z-w)^2},\cr
{1\over 2}\corr{J(z)\overline{J}(w)+\overline{J}(z)J(w)}&=
{S(\tau)\over(z-w)(\overline{z}-\overline{w})},\cr}
\neq \label\defrs $$
where $\tau=\ln (z-w)(\overline{z}-\overline{w})\Lambda^2$, and
$\Lambda$
is the renormalisation scale of the theory. For
reflection-positive theories,
$R(\tau)$ is a positive function while $S(\tau)$ is negative.
AT CFT, $S(\tau)=0$ and $R(\tau)$ becomes a
constant, which we shall denote by $K_{CFT}$. In the case of a
non-abelian
conserved gauge current, it is related to the level of the central
extension of the corresponding affine Lie algebra \KZ.

Among the two current Ward Identities of $J_\alpha$, we have
$$ \eqalign{
\corr{\left(\partial_{\overline{z}}J(z)+\partial_z\overline{J}(z)
\right)
J(w)}&=0,\cr
\corr{J(z)
\left(\partial_{\overline{w}}J(w)+\partial_w\overline{J}(w)\right
)}&=0.\cr}
\neq \label\csrveq $$
Using definitions \defrs\ and subtracting the two equations in
\csrveq\ we obtain
$$
\dot R(\tau)+\dot S(\tau) = S(\tau),
\neq $$
where $\dot R = {d\over d\tau} R$. From that,
a quantity decreasing along the RG flow can immediately be
defined,
namely,
$$\eqalign{
K(\tau)&=2(R(\tau)+S(\tau)),\cr
\dot K(\tau)&=2S(\tau)\leq 0.\cr}
\neq $$
At a fixed point, since $S(\tau)=0$, $K(\tau)$ coincides with the
constant
coefficient of the two point function and is stationary.
If we fix the value of $\tau$ (say, to $\tau_0$), $K(\tau=\tau_0)$
will depend only
on the coupling constants of the theory, $g=\{ g_i\}$. Then, the
RG flow is given by
$$
\beta_i (g) {\partial\over\partial g_i} K(g) =
{1\over 2}(z-w)(\overline{z}-\overline{w})
\corr{J(z)\overline{J}(w)+\overline{J}(z)J(w)}\vert_{\tau=\tau_0}.
\neq \label\kbetaflow $$

A sum rule {\it \`a la Cardy} \C\ can immediately be derived from
the previous result. In effect, we can rewrite
$$
\dot K = {d\over d\tau} K = r^2 {d\over dr^2} K,
\neq \label\kreq $$
where $r^2=(z-w)(\overline{z}-\overline{w})$. The total change in
the coefficient $K_{CFT}=K(\tau)|_{CFT}$ is given by
$$\eqalign{
\Delta K_{CFT} \equiv
K_{UV}-K_{IR}&=K(\tau=-\infty)-K(\tau=\infty)\cr
&=K(r^2=0)-K(r^2=\infty)=-\int_0^\infty dr^2\ {d\over dr^2}
K,\cr}
\neq $$
which, in the light of equation \kreq, becomes
$$
\Delta K_{CFT}=-{2\over \pi}\int d^2x\ {1\over 2}
\corr{J(z)\overline{J}(w)+\overline{J}(z)J(w)}.
\neq \label\fstsr $$

We have so established the irreversibility of the RG flow for
theories
with conserved vector currents, using arguments similar to
Zamolodchikov's in his original proof of the $c$-theorem.

\section{$k$-theorem and spectral representation}
\parag{Spectral decomposition of a two current correlation
function}
Just like the $c$-theorem, the $k$-theorem also admits a proof
using the
spectral decomposition of a two point function \F\ \CFL. Actually,
this is the original way the theorem was demonstrated in reference \F.
Before going on with the proof, we shall study some necessary properties
of the correlation function of two conserved vector currents.

Consider again a two-dimensional Quantum Field Theory which is
Lorentz
invariant and unitary. Let $J_\alpha (x)$ be a conserved current
in the
theory, at classical and quantum level. Let us study the two
current
correlation function. By inserting a resolution of the identity
made out
of representations of the Poincar\'e group, we obtain the
spectral
decomposition of $\corr{J_\alpha (x) J_\beta (0)}$ (see, for
instance, \IZ),
$$
\corr{J_\alpha (x) J_\beta (0)} = \left(
\partial_\alpha\partial_\beta
- \eta_{\alpha\beta} \sq\right) {1\over\pi}
\int_0^\infty d\mu\ k(\mu)\Delta(x,\mu),
\neq \label\xssr $$
being $\Delta(x,\mu)$ the free propagator for a spinless particle
of mass $\mu$, namely, in two dimensions,
$$
\Delta(x,\mu)={1\over 2\pi} K_0(\mu \norm{x}).
\neq $$
The $K_n$ symbol stands, as usual, for the modified Bessel
function of
order $n$. The spectral density $k(\mu)d\mu$ measures the density
of degrees of
freedom coupling to the current at distance $\mu^{-1}$. It is
made
out of objects living in the Hilbert space of the theory, being
therefore well defined.

In complex coordinates, with the conventions established in the
previous section, equation \xssr\ becomes
$$\eqalign{
\corr{J(z)J(w)}&={1\over8\pi^2}{\overline{z}\over z}\int_0^\infty
d\mu\
k(\mu)\mu^2K_2(\mu\norm{x}),\cr
{1\over 2}\corr{J(z)\overline{J}(w)+\overline{J}(z)J(w)}&=
-{1\over8\pi^2}\int_0^\infty d\mu\ k(\mu)\mu^2K_0(\mu\norm{x}) +
{1\over4\pi}\delta^2(x)\int_0^\infty d\mu\ k(\mu).\cr}
\neq \label\ccdecomp $$

To determine the spectral density, one can proceed in several
ways, from
using form factors to direct calculation \FLV. However, equation
\xssr\ written in momentum space can be recast into a
dispersion relation, which provides a very convenient
way to compute $k(\mu)$. In effect, in Euclidean space, we have,
$$
\Pi_{\alpha\beta}(p) = \left( p_\alpha p_\beta -
\delta_{\alpha\beta} p^2\right) {1\over\pi}
\int_0^\infty d\mu\ k(\mu) {1\over p^2+\mu^2}.
\neq \label\pssr $$
In the following, $\Pi_{\alpha\beta}(x)$ shall denote the
two current
correlation function $\corr{J_\alpha (x) J_\beta (0)}$. If we
take the trace over both sides of equation \pssr, we have
$$
\delta^{\alpha\beta}\Pi_{\alpha\beta}(p) ={1\over\pi}
\int_0^\infty d\mu\ k(\mu) {\mu^2\over p^2+\mu^2} + {\rm
constant},
\neq \label\drel $$
which can be interpreted as a dispersion relation.
Therefore, $k(\mu)$
is related to the imaginary part of the trace of
$\Pi_{\alpha\beta}$,
$$
k(\mu)={2\over\mu}\ {\rm Im\,}
\delta^{\alpha\beta}\Pi_{\alpha\beta}
(p^2=-\mu^2).
\neq \label\disprel $$
This provides a suitable formula to evaluate the spectral
density.

Yet preserving full generality, one can make some statements
regarding the functional form of the spectral density \F\ \CFL.
One has to start
by looking at CFT, where no scales are present. By power counting
arguments, it can be established that only two behaviours are
allowed,
$$\eqalign{ {\rm (i)\ } k_{CFT}(\mu)&= k_0\delta(\mu),\cr
{\rm (ii)\ } k_{CFT}(\mu)&= {k_0\over\mu},\cr} $$
being $k_0$ a constant related to the constant
coefficient of
the two current correlation function. The case (ii) rises
non-existing
IR singularities which leaves form (i) as the correct one.
Then, $\Pi_{\alpha\beta}(x)$ at CFT is
$$
\Pi_{\alpha\beta}(x)\vert_{CFT}={k_0\over2\pi^2}\left(
2{x_\alpha x_\beta\over x^2}-\delta_{\alpha\beta}\right){1\over
x^2},
\neq $$
in agreement with reference \OP.
For a general theory, the spectral function should then
have the form,
$$
k(\mu)= k_0\delta(\mu) + k_1(\mu,\Lambda),
\neq \label\kkk $$
being $\Lambda$ the renormalisation scale in the theory and
$k_1(\mu,\Lambda)$ a smooth function, non singular when $\mu\to0$.
The
delta term will account for the contribution of the massless
degrees of
freedom coupling to the current while $k_1(\mu,\Lambda)$ comes
from the intermediate states of mass $\mu>0$. The role of
unitarity
is to ensure the positivity of $k_0$ and $k_1(\mu,\Lambda)$.

\parag{Spectral proof of the $k$-theorem}
The considerations
about the functional form of $k(\mu)$ are the first step
towards the
spectral proof of the $k$-theorem. In order to finish the
demonstration after reference \F\ we shall
study the short and long distance behaviour of the two
current correlation function. At short distances, when $x\to
0$, the two current correlation function \xssr\ takes the form of
that at CFT,
$$
x\to 0 \Rightarrow \Pi_{\alpha\beta}(x)\to {1\over 2\pi^2}\left(
2{x_\alpha
x_\beta\over x^2}-\delta_{\alpha\beta}\right){1\over
x^2}\int_0^\infty
d\mu\ k(\mu).
\neq $$
This expression recovers a more familiar aspect in complex
coordinates,
$$
x\to 0 \Rightarrow \cases{ \corr{J(z)J(w)}\to {\displaystyle
{1\over\pi^2} {1\over z^2}\int_0^\infty d\mu\ k(\mu)},&\cr
{1\over 2}\corr{J(z)\overline{J}(w)+\overline{J}(z)J(w)}\to
0.&\cr}
\neq $$
Identically, in the long distance limit, $x\to\infty$,
$\Pi_{\alpha\beta}$ has the CFT form. This time,
$$
x\to\infty\Rightarrow \cases{\corr{J(z)J(w)}\to {\displaystyle
{1\over \pi^2}
{1\over z^2}\lim_{\eps\to 0}\int_0^\eps d\mu\ k(\mu)},&\cr
{1\over 2}\corr{J(z)\overline{J}(w)+\overline{J}(z)J(w)}\to
0.&\cr}
\neq $$
We can identify the UV and IR coefficients of the two point
function,
$$
\eqalign{ k_{UV} &=\int_0^\infty d\mu\ k(\mu),\cr
k_{IR} &=\lim_{\eps\to 0}\int_0^\eps d\mu\ k(\mu),\cr}
\neq $$
{}From the general functional form for $k(\mu)$ \kkk, we get
$$
k_{UV}=\int_0^\infty d\mu\ k(\mu) = k_{IR} + \int_0^\infty d\mu\
k_1(\mu,\Lambda).
\neq $$
For unitary theories, by positivity of the spectral density, we
have
$$
k_{UV}\geq k_{IR}.
\neq \label\kuvkir $$
We have thus shown that the coefficient of the two current
correlation function at CFT decreases along the RG flow.
The RG flow will thus be irreversible.
The amount the coefficient decreases is given
by the spectral sum rule
$$
\Delta k=k_{UV}-k_{IR} = \int_0^\infty d\mu\ k_1(\mu,\Lambda).
\neq $$
This sum rule deserves some considerations.
Being $J_\alpha(x)$ a conserved current, the spectral density
$k(\mu)d\mu$ gets no renormalisation. Therefore, a change of
scale is absorbed as
$$
k_\lambda(\mu)d\mu = k(\lambda\mu)\lambda d\mu
\neq $$
and the sum rule
$$\eqalign{
\Delta k=\int_0^\infty d\mu\ k_1(\mu,\Lambda)&= \int_0^\infty
\lambda d\mu\ k_1(\lambda\mu,\Lambda)\cr &=\int_0^\infty d\mu\
k_1(\mu,
{\Lambda\over\lambda})\cr}
\neq $$
can be computed at any point along the RG flow, that is, at any
$\Lambda$. The UV limit corresponds to $\lambda\to\infty$, which
is
equivalent to setting to zero the scale $\Lambda$. In this limit
the spectral function must vanish for all $\mu\not= 0$ but still
have a
finite integral, which means that it becomes a representation of
a Dirac
$\delta$-function. This is the rigorous expression of the picture
one
draws for the behaviour of the spectral density along the RG flow
: when we perturb the initial CFT some degrees of freedom coupling
to the
current become massive eventually decoupling in the IR limit. If
the
relevant perturbation does not succeed to turn massive any of the
degrees of freedom in the theory, the spectral density will
maintain the
original UV delta form, and we will recover the equal sign in
equation
\kuvkir.

With these tools one can
build a function which decreases along the RG flow and
which is stationary at fixed points in order to contact with the
proof
in the previous section. For this purpose,
we just need a positive smooth function $f(\mu)$ such that
$f(0)=1$, $f$
decreases exponentially for large $\mu$ and its derivative is
negative.
{}From the spectral density we have
$$\eqalign{
k(\Lambda)=\int_0^\infty d\mu\ k(\mu)f(\mu)&= k_{IR} +
\int_0^\infty
d\mu\ k_1(\mu,\Lambda)f(\mu),\cr
\Longrightarrow\quad\Lambda {d\over d\Lambda} k(\Lambda) &\leq
0.\cr}
\neq $$
Since the full dependence of $k(\Lambda)$ in $\Lambda$ will be
given by
its dependence in the couplings of the theory, the previous
equation
becomes equivalent to equation \kbetaflow. The function $K(g)$ in
the
previous section would correspond to a particular choice of the
function
$f(\mu)$.

\parag{Sum rules}
Sum rules are used to evaluate quantitatively the change in $k_0$
along the flow. We have just seen how the spectral decomposition
formulation of
the $k$-theorem provides immediately a manner to compute such
change, $\Delta k=k_{UV}-k_{IR}$, namely,
$$
\Delta k=\int_0^\infty d\mu\ k_1(\mu,\Lambda)=
\lim_{\eps\to 0}\int_\eps^\infty d\mu\ k(\mu).
\neq \label\zthsr $$
Besides, this spectral proof allows to recover the sum rule
\fstsr\ of
the previous section. We just have to use the decomposition
\ccdecomp\
to see that
$$\eqalign{
\int d^2x\
{1\over 2}\corr{J(z)\overline{J}(0)+\overline{J}(z)J(0)} =&
-{\pi\over 4}\int_0^\infty d\mu\ k(\mu)\cr
&+ {\rm contribution\ of\ the\ contact\ term}.\cr}
\neq $$
Neglecting the contribution of the contact term, we recover
equation \fstsr,
$$
\Delta k=\lim_{\eps\to 0}\int_\eps^\infty d\mu\ k(\mu)=
-\lim_{\eps\to 0}\int_{\norm{x}>\eps} d^2x\
{2\over\pi}\corr{J(z)\overline{J}(0)+\overline{J}(z)J(0)}.
\neq \label\sndsr $$
Recall that the coefficients $k$ and $K_{CFT}$ are related by a
factor $4$.

Similarly, a third sum rule can be found using $\corr{J(z)J(0)},$
$$
\Delta k=\lim_{\eps\to 0}\int_\eps^\infty d\mu\ k(\mu)=
-{4\over\pi}\lim_{\eps\to 0}\int_{\norm{x}>\eps} d^2x\
z{\partial\over\partial\overline{z}}\corr{J(z)J(0)}.
\neq \label\trdsr $$

Thus, given a theory, we are armed with several tools to compute
the change in the constant coefficient of the two point function.

\section{Example : fermions in two dimensions}
In order to illustrate the features appearing in the previous
sections,
we shall study the case of two dimensional free fermions.

\parag{The Abelian gauge current}
Let $\psi(x)$ describe a free two-dimensional Dirac fermion of
mass $m$.
The theory has a well known U(1) symmetry whose conserved current
is
$$
J_\alpha (x) = :\overline{\psi}\gamma_\alpha\psi:(x)
\neq $$
The two current correlation function, $\Pi_{\alpha\beta}$,
happens to be the
one-loop contribution to the vacuum polarisation of the photon in
the
Schwinger model and encodes the chiral anomaly. If the mass is
set to zero, we recover a Conformal Field Theory. Then,
$$
\Pi_{\alpha\beta}(x)\vert_{m=0}={1\over2\pi^2}\left(
2{x_\alpha x_\beta\over x^2}-\delta_{\alpha\beta}\right){1\over
x^2},
\neq $$
which is the standard form for a two vector conserved current
correlation function in CFT. This implies that $k_0=1$. Actually, the
normalisation of equation \xssr\ was devised to fit this result. When we
switch on the massive perturbation, the theory flows towards the
trivial fixed point where no degree of freedom couples to the current
and, thus $k_{IR}=0$. So, since $k_{UV}=1$, we expect $\Delta
k=1$. In order to check this prediction, we need the massive
version of $\Pi_{\alpha\beta}$. In momentum space, we have
$$
\Pi_{\alpha\beta}(p) = \left(\delta_{\alpha\beta}- {p_\alpha
p_\beta\over p^2}\right)\left[ {m^2\over\pi}\int_0^1dt\ {1\over
-t^2p^2
+tp^2+m^2}-{1\over\pi}\right].
\neq \label\dfpab $$
{}From this equation, we can immediately check, using the sum rule
\sndsr, that, in effect,
$$
\Delta k = 1.
\neq $$

We can also use equation \dfpab, together with the dispersion
relation \disprel, to evaluate the spectral density of
$\Pi_{\alpha\beta}$. We find
$$
k(\mu,m)={4m^2\over\mu^2\sqrt{\mu^2-4m^2}}\theta(\mu^2
-4m^2),
\neq \label\tdfk $$
being $\theta(x)$ the step function. This result coincides with the
computation in reference \FS. Note the two particle production
threshold appearing. By the field content of the current,
when introducing a resolution of the identity made out of
Poincar\'e representations, one can
see that only two particle states will saturate the correlation
function. Precisely, Poincar\'e invariance will not allow any two
particle state below the mass gap, so the threshold is a feature
that
the correct solution should exhibit. With equation \tdfk, we can
immediately check that
$$ \Delta k = \int_0^\infty d\mu\ k(\mu,m)= 1.
\neq $$
In the limit $m\to\infty$, $k(\mu,m)$ tends to zero as expected.
On the other end, in the limit $m\to 0$, $k(\mu,m)$ reproduces the
behaviour of
a Dirac delta function, with coefficient 1, recovering so the CFT
behaviour. This fact can be checked by integrating $k(\mu,m)$
with a test function $f(\mu)$. In the integral, one must
rescale the integration variable
$\mu$ by $m$ in order to expand $f$ in Taylor series. The
non-vanishing terms in the expansion when $m\to0$ should be kept to
compare the result with that of the action of a delta function. The
spectral decomposition is, thus, as announced in section 3, a
representation of
the Dirac delta function when the scale of the theory is removed.

\parag{Non abelian gauge currents}
Consider now a theory of two-dimensional free fermions, $\psi_i$,
of
equal mass $m$, with
a non-abelian symmetry group, SU(N) for instance. Let
$\Psi$ denote the $N$-plet of fermions and $t^a$ the generators
of the
algebra of the symmetry group in the representation in which the
fermions live. Then the non-abelian conserved current is
$$
J^a_\alpha(x)=:\overline\Psi t^a\gamma_\alpha\Psi:(x)
\neq $$
The two current correlation function is rapidly related to
the Dirac fermion one by the formula
$$
\Pi^{ab}_{\alpha\beta}(x)=\corr{J^a_\alpha(x)J^b_\beta(0)}= C_A
\delta^{ab} \Pi_{\alpha\beta}(x),
\neq $$
where ${\rm Tr\,} t^at^b = C_A\delta^{ab}$. From this, we
deduce that $k_0=C_A$ when $m=0$ and therefore, when the
mass perturbation is switched on, $\Delta k=C_A$. We
can change the normalisation of the generators,
$$\eqalign{
\widehat{J}^a_\alpha(x)&=:\overline\Psi
\widehat{t}^a\gamma_\alpha\Psi:(x),\cr
{\rm Tr\,} \widehat{t}^a\widehat{t}^b&=\delta^{ab},\cr}
$$
in order to have $k_0=1$. If we now consider $n$ copies of the
theory, this is equivalent to a Wess-Zumino-Novikov-Witten
model of level $k=n$ \W. When we add a mass to one of
the copies, the
total theory will change from $k_{UV}=n$ to $k_{IR}=n-1$,
$k$ decreasing along the RG flow. This is a particular
case of
theory with an affine Lie algebra associated in which we check
the decreasing of the level of the algebra along the flow.
\vfill\eject
\section{Extension to higher dimensions}
\parag{Spectral decomposition in more than two dimensions}
The extension to more than two dimensions of the previous results
on the
irreversibility of RG flows for theories with vector conserved
currents would be very interesting. To achieve this goal, the
spectral representation approach seems most promising because
its essential set up
is made irrespective of the dimension of space time.
On the other hand, the proof
{\it \`a la Zamolodchikov} is very much constrained by to the two
dimensional formalism. Besides,
the vector current case looks a priori simpler than the
$c$-theorem one. In effect, the spectral decomposition of the two
stress tensors correlation function has two spin structures \CFL,
while the two current correlation function has only
one. For this reason we shall study the extension of the
$k$-theorem to higher dimensions through this approach.

 We consider this time a $d$-dimensional Quantum Field Theory
which is Lorentz invariant and unitary.
We denote by $J_\alpha (x)$ the required
conserved current at classical and quantum level.
The analysis of the spectral decomposition of $\corr{J_\alpha (x)
J_\beta (0)}$ is essentially the same than the two dimensional case.
We have
$$
\Pi_{\alpha\beta}(x)=
\corr{J_\alpha (x) J_\beta (0)} = \left(
\partial_\alpha\partial_\beta
- \eta_{\alpha\beta} \sq\right) \int_0^\infty d\mu\ k(\mu)
\Delta(x,\mu),
\neq \label\dxssr $$
being $\Delta(x,\mu)$ the $d$-dimensional free propagator for a
spinless particle of mass $\mu$,
$$
\Delta(x,\mu)={1\over 2\pi}\left({\mu\over2\pi
x}\right)^{d-2\over2}
K_{d-2\over2}(\mu \norm{x}).
\neq $$
By rewriting equation \dxssr\ in Euclidean momentum space and
taking the
trace over space time indices, we obtain, just like in the
two-dimensional case, a dispersion relation. It relates the
spectral density $k(\mu)$ with the imaginary part of the
correlation function,
$$
k(\mu)={2\over (d-1)\pi\mu}\ {\rm Im\,}
\delta^{\alpha\beta}\Pi_{\alpha\beta} (p^2=-\mu^2).
\neq \label\ddisprel $$
For unitary theories, $k(\mu)$ is positive, as in the two
dimensional case.

In more than two dimensions it becomes harder to put forward a
general statement on the form of $k(\mu)$, in the line of
equation \kkk. If we start by the form at CFT,
it is well known that, in any dimension, Conformal Invariance
completely
constrains the form of two point functions (see, for instance \G\
and \OP).
In the case of conserved vector currents, this compulsory
behaviour is
$$
\corr{J_\alpha(x) J_\beta(0)} = {A\over
\norm{x}^{2(d-1)}}\left(
2{x_\alpha x_\beta\over x^2}-\delta_{\alpha\beta}\right),
\neq \label\dcftb $$
where $A$ is a constant.
Regarding the functional form of $k(\mu)$, which must reproduce
equation
\dcftb, dimensional analysis allows only two possibilities,
$$ \eqalign{{\rm (i)\ } k_{CFT}(\mu)&=
k_0\mu^{d-2}\delta(\mu),\cr
{\rm (ii)\ } k_{CFT}(\mu)&= k_0\mu^{d-3},\cr} $$
with, $k_0$, of course, a constant. In contradistinction to
two dimensions, case (ii) does not arise IR singularities,
and it
is perfectly well defined. Actually, if we plug case (i) into
formula
\dxssr, the two current correlation function vanishes (which
implies that
$A=0$). However, form (i) allows the splitting of a generic
spectral density into
$$k(\mu)=k_0\mu^{d-2}\delta(\mu)+k_1(\mu,\Lambda),$$ being
$k_1(\mu,\Lambda)$ smooth in the limit $\mu\to 0$.
In this way, the
contribution from massless degrees of freedom is separated from
the
contribution of massive ones, just like in the two dimensional
analysis.
On the other hand, case (ii) delivers the general result \dcftb.
The constants $k_0$ and $A$ are then related by the following
equation,
$$
k_0={2^{4-d}\pi^{d/2}\over\Gamma(d)\Gamma({d\over2}-1)}A,
\neq \label\kk $$
where $\Gamma(t)$ denotes the Euler $\Gamma$-function. The
splitting of
$k(\mu)$ makes no sense any more since for such kind of currents,
massless degrees of freedom couple to all distances $\mu^{-1}$
and thus they are hard to separate from massive ones.

\parag{Examples : free bosons and fermions}
To show how the spectral decomposition works in more than two
dimensions, we shall consider two very simple examples. These are
free bosons and fermions of mass $m$, with a U(1) symmetry.

Let us start by studying a free complex boson $\ph (x)$ in $d$
dimensions. The theory has a U(1) conserved current
$$
J^{\rm bos}_\alpha (x) = {\rm i} :\ph^* {\buildrel
\leftrightarrow \over
\partial_\alpha} \ph : (x).
\neq $$
A first test consists in checking the CFT behaviour of the two
current correlation function for the $m=0$ case. One recovers the
form \dcftb\
with
$$
A^{\rm bos}={2\over (d-2)S_d^2},
\neq \label\bdcftk $$
where $S_d=2\pi^{d/2}/\Gamma(d/2)$. Since the value of $A$ is
non-vanishing, we expect that the spectral density will follow
the power
law behaviour rather than the delta one. When switching on the
massive perturbation, the computation of the two current
correlation function in Euclidean momentum space delivers
$$
\Pi_{\alpha\beta}^{\rm bos}(p)=(p_\alpha p_\beta -
\delta_{\alpha\beta} p^2){\Gamma\left(2-{d\over2}\right)\over
(4\pi)^{d\over2}}\int_0^1 dt\ {(1-2t)^2\over\left(t(1-t)p^2+m^2
\right)^{2-{d\over2}}},
\neq $$
Using the dispersion relation \ddisprel, we retrieve the spectral
density of the two current correlation function. We find
$$
k^{\rm bos}(\mu)={1\over2(4\pi)^{d-1\over2}}{1\over\Gamma
\left({d+1\over2}\right)}
\left({\mu\over2}\right)^{d-3}
\left(1-{4m^2\over\mu^2}\right)^{d-1\over
2}\theta(\mu^2-4m^2).
\neq $$
We see the two particle production threshold appearing as a
healthy sign
of the result. Besides, if we take the limit $m\to0$, we recover
the
expected power law behaviour and the coefficient $k_0$ matches
the conformal result \bdcftk. Another interesting limit is the
$d\to2$ one.
In effect, we see that the power law behaviour still holds when
we
further take $m\to 0$. This is due to the IR troubles of the free
massless boson in two dimensions. The generalisation of these
results
to the non-abelian case is straightforward. If $t^a$ are the
generators
of the non-abelian symmetry group of a free boson theory, with
the normalisation,
$$\eqalign{
J_\alpha^a (x)&= :\ph_i t_{ij}^a\partial_\alpha \ph_j:(x),\cr
{\rm Tr\ }(t^at^b)&=-N_\ph\delta^{ab},\cr}
\neq $$
we have,
$$
\corr{J_\alpha^a(x)J_\beta^b(0)} =
{1\over2}N_\ph\delta^{ab}\Pi_{\alpha\beta}^{\rm bos}(x).
\neq $$
In the massless limit, we recover the results from reference \OP.

The second example is a free $d$-dimensional fermion $\psi(x)$ of
mass $m$. The U(1) conserved current is
$$
J_\alpha(x)=:\overline{\psi}\gamma_\alpha\psi:(x).
\neq $$
Again we start by checking the CFT behaviour in the $m=0$ case.
We have
$$
A^{\rm fer}={2^{d\over 2}\over S_d^2}.
\neq \label\fdcftk $$
Then,
$$
\Pi_{\alpha\beta}^{\rm fer}(p)=-2(p_\alpha p_\beta -
\delta_{\alpha\beta} p^2){\Gamma\left(2-{d\over2}\right)\over
(2\pi)^{d\over2}}\int_0^1 dt\ {t(1-t)\over\left(t(1-t)p^2+m^2
\right)^{2-{d\over2}}},
\neq $$
and
$$
k^{\rm fer}(\mu)={\sqrt{\pi}\over(2\pi)^{d\over2}\Gamma
\left({d-1\over2}\right)}\left({\mu\over2}\right)^{d-3}
\left(1-{4m^2\over\mu^2}\right)^{d-3\over2}\left[ 1- {1\over
d-1}\left(1-{4m^2\over\mu^2}\right)\right]\theta(\mu^2-4m^2).
\neq $$
Just like for the boson case, the two particle production
threshold is found. The limit $m\to0$ also delivers the expected
power
law behaviour with a coefficient compatible with \fdcftk. For the
$d\to
2$ limit, we recover equation \tdfk, and the following limit
$m\to 0$
gives the correct delta result. Curiously, we can invert the
order of
the two limits (first $m\to0$ and then $d\to2$), and see how the
delta
behaviour is surprisingly recovered from a clever appearing of
$(d-2)$
factors. Regarding the non-abelian generalisation, we just take
$$\eqalign{
J_\alpha^a (x)&= :\overline{\psi}_i t_{ij}^a\gamma_\alpha
\psi_j:(x),\cr {\rm Tr\ }(t^at^b)&=-N_\psi\delta^{ab},\cr}
\neq $$
to get
$$
\corr{J_\alpha^a(x)J_\beta^b(0)} =
N_\psi\delta^{ab}\Pi_{\alpha\beta}^{\rm fermion}(x),
\neq $$
which is again consistent with the results from reference \OP.

\parag{Extension of the $k$-theorem to higher dimensions}
The two possible behaviours for the spectral density $k(\mu)$
described in paragraph 5.{\oldstyle 1} are the same that
exhibit, respectively, the
spin zero and the spin two spectral densities, when we
decompose the correlation function of two stress tensors for
$d>2$. Therefore, we shall
briefly sketch the discussion in references \CFL\ and \CLV\ about
whether a quantity decreasing along the RG flow exists.

In case (i), such a quantity can be built up by following closely
the two dimensional argument. This can be done because the
contribution of
massless degrees of freedom is clearly separated from that of the
massive ones. However, at present, it is not known
whether the coefficient $k_0$ can be defined always and uniquely
for any
CFT, like $c$ or $k$ are in two dimensions.
Then, we cannot speak about irreversibility of the RG flow
since a theory could come back to herself in a RG loop, with a
different
value of $k_0$. The only existing approach defines $k_0$ using a
limiting procedure away from criticality, which is only
consistent if
the space of theories is a manifold. In this case, we can
rigorously
speak about irreversibility of the RG flow. Therefore,
we are in the same situation than in the
$c$-theorem with the spin zero spectral density.

In case (ii), on the other hand, the constant $k_0$ is well
defined at
the Conformal point, by the two point function, following formula
\kk. However, a quantity decreasing along the RG flow cannot
immediately be
built since, despite $k(\mu)$ is positive, information about the
positivity of the derivative of $k(\mu)$ would be needed. Such
information requires dynamical considerations, spoiling a general
statement about the RG flow. This is precisely the same problem
encountered in the analysis of the spin two spectral density of
the two
stress tensor correlation function. In the case of the examples
in paragraph
5.{\oldstyle 2}, both theories flow from the gaussian Conformal
point
$m=0$ towards the trivial fixed point. This means that $k_0$ goes
from
the free theory value towards 0, effectively decreasing along the
flow.
However, this is only a very simple example, and does not allow
to draw any conclusion for more complicated flows in interacting
theories.

\bigskip
Summing up, we have shown the
irreversibility of RG flows in two dimensions for theories with
vector
conserved currents. The naive extension of the spectral
version
of the $k$-theorem encounters the same difficulties than its
corresponding version of the $c$-theorem, in spite of the a
priori
simplification drawn by the only spin structure of the
decomposition of
the correlation function of two conserved currents. From here,
one can
look for particular
applications of the two dimensional result to a number of
conserved currents. Regarding the extension to higher
dimensions, the study of particular flows in interactive theories
might
shed some light on the behaviour of the derivative of $k(\mu)$ in
case
(ii), just to see if a theorem is ruled out by a counterexample
or it
is verified in some specific cases. However, these extensions are
beyond the scope of this introductory paper.

\vfill
{\flrm Acknowledgements}\par
This work has been partially supported by the Ministerio de
Educaci\'on y Ciencia through an FPI grant, CICYT under contract AEN
90-0033 and NATO under contract CRG-910890.
I would like to acknowledge the warm hospitality of the Lawrence
Berkeley Laboratory where part of this work was developped.
I thank very specially Jos\'e I. Latorre for introducing
me to the subject and sharing his insight.
I am indebted to O.Alvarez, P.E.Haagensen and J.Soto
for discussions, and A.Cappelli for his comments on the
manuscript.

\immediate\closeout\rfile
\vfill \eject
\centerline{\flrm References} \bigskip
\input refs.aux
\vfill \eject
\bye